\documentclass[12pt,twoside,epsf,epsfig]{article}
\usepackage{fleqn,espcrc1}

\usepackage{epsfig}
\usepackage[figuresright]{rotating}

\def\beq{\begin{equation}}
\def\eeq{\end{equation}}
\def\bea{\begin{eqnarray}}
\def\eea{\end{eqnarray}}
\def\bq{\begin{quote}}
\def\eq{\end{quote}}

\def\gappeq{\mathrel{\rlap {\raise.5ex\hbox{$>$}}
{\lower.5ex\hbox{$\sim$}}}}

\def\lappeq{\mathrel{\rlap{\raise.5ex\hbox{$<$}}
{\lower.5ex\hbox{$\sim$}}}}

\def\Toprel#1\over#2{\mathrel{\mathop{#2}\limits^{#1}}}
\def\FF{\Toprel{\hbox{$\scriptscriptstyle(-)$}}\over{\nu}}

\newcommand{\AmS}{{\protect\the\textfont2
  A\kern-.1667em\lower.5ex\hbox{M}\kern-.125emS}}

\title{Strangeness and Hadron Structure}

\author{John Ellis\address{Theoretical Physics Division, CERN \\ 
        CH - 1211 Geneva 23}}%
       
\begin{document}

\maketitle

\begin{abstract}
The nucleon wave
function may contain a significant component of $\bar s s$
pairs, according to several measurements including the $\pi$-nucleon
$\sigma$ term, charm production and polarization effects in 
deep-inelastic scattering. In addition, there are excesses of
$\phi$ production in LEAR and other experiments,
above predictions based the naive Okubo-Zweig-Iizuka rule, 
that may be explained if the nucleon wave function contains a
polarized ${\bar s}s$ component. This model also reproduces
qualitatively data on $\Lambda$ polarization in deep-inelastic
neutrino scattering. The strange component of the proton is
potentially important for other physics, such as the search for 
astrophysical dark matter.
\end{abstract}

\begin{center}
CERN-TH/2000-112 $\;\;\;\;\;$ hep-ph/0005322
\end{center}

\section{Strange Ideas}

Does the nucleon wave function contain a (large) strange component?  The
starting point for any discussion of the quark flavour content is the amazingly
successful na\"\i ve quark model (NQM), in which $\vert p> = \vert UUD>$, with
each constituent quark weighing $m_{U,D} \sim 300$~MeV~\cite{NQM}.  A
simple
non-relativistic $S$-wave function with $v/c \ll 1$ is surprisingly successful: 
even better is a simple harmonic oscillator potential with a $D$-wave admixture
of 6\% (in amplitude)~\cite{IK}.  For comparison, we recall that the
deuteron and $^3He$
wave functions contain similar $D$-wave admixtures.  Neglecting any such
$D$-wave component, the (overly?) na\"\i ve quark model would predict that the
proton spin is the algebraic sum of the constituent quark spins: 
$\underline{s}_P
= \underline{s}_U + \underline{s}_U  + \underline{s}_D$.  Axial current matrix
elements indicated that the quark spins might contribute at most 60\% of the
proton spin, even before the EMC and its successor
experiments~\cite{oldspin}, but we return to
this later.  Whatever the partial-wave decomposition, if the proton only
contains $\vert UUD\rangle$ Fock states, and one neglects pair creation, 
a consequence is the
Okubo-Zweig-Iizuka (OZI) rule~\cite{OZI} forbidding the coupling of the
proton $\vert \bar
s s\rangle$ mesons.  The validity of the OZI rule is another
major theme of this talk.

Although the NQM is very successful, it has never been derived from QCD, and is
expected to be wrong and/or incomplete~\cite{NQM}.  The validity of chiral
symmetry informs
us that the light quarks are indeed very light:  $m_{u,d} < 10$~MeV, $m_s \sim
100$~MeV~\cite{chisym}.  These estimates refer to the current quarks
visible in short-distance
or light-cone physics.  Such current quarks should be relativistic:  $v/c \sim
1$, and there is no obvious reason why pair production of $\bar u u$, $\bar d d$
or $\bar s s$ should be suppressed.  Indeed, non-perturbative interactions 
with the
flavour content $(\bar u u)(\bar d d)(\bar s s)$ are believed to be
important~\cite{instantons}
and the light quarks are known to condense in the vacuum~\cite{chisym}:
\beq
\langle 0\vert \bar u u\vert 0 \rangle \simeq \langle 0 \vert \bar d d \vert 0
\rangle \equiv \langle 0 \vert \bar q q \vert 0 \rangle \sim \langle 0 \vert
\bar s s \vert 0\rangle \sim \Lambda^3_{QCD}~,
\label{one}
\eeq
where
\beq
m^2_\pi \simeq \frac{m_u + m_d}{f^2_\pi}~\langle 0\vert \bar q q \vert 0
\rangle~,~~m^2_K \simeq \frac{m_s}{f^2_K}~\langle 0 \vert \bar s s \vert 0
\rangle~.
\label{two}
\eeq
Inside a proton or other hadron, one would expect the 
introduction of colour charges to perturb the ambient vacuum.  Since
the connected matrix element
\beq
\langle p \vert \bar q q \vert p \rangle \equiv \langle 
\bar q q \rangle_{\rm full} -
\langle 0 \vert \bar q q \vert 0 \rangle~~ \langle p\vert p\rangle~,
\label{three}
\eeq
one could expect that $\langle p \vert \bar s s \vert p \rangle \not= 0$.

There are many suggestions for improving the NQM, such as bag
models~\cite{bag} -- in which
relativistic quarks are confined within a cavity in the vacuum, 
chiral solitons~\cite{Skyrme} --
which treat nucleons as coherent mesonic waves, and hybrid
models~\cite{hybrid} -- which place
quarks in cavities inside mesonic solitons.  
As an example of the opposite extreme
from the NQM, consider the Skyrme model.

In this model~\cite{Skyrme}, the proton is regarded as a solition 'lump'
of meson fields:
\beq
\vert p \rangle = V(t)U(\underline x)V^{-1}(t)
\label{four}
\eeq
with
\beq
U(\underline x) = {\rm exp} \left( \frac{2\pi i \underline\tau \cdot
\underline\pi(\underline x)}{f_\pi} \right)~,
\label{five}
\eeq
where $\underline \pi(\underline x)$ are SU(2) meson fields and 
the $\underline\tau$ are isospin matrices, 
and $V(t)$ is a time-dependent rotation matrix in
both internal SU(3) space and external space.  The Skyrme model embodies chiral
symmetry, and is justified in QCD when the quarks are very light:  $m_q \ll
\Lambda_{QCD}$, which is certainly true for $q = u, d$, but more debatable for
the strange quark.  The Skyrme model should be good for long-distance
(low-momentum-transfer) properties of nucleons, such as (the ratios of) magnetic
moments $\mu_{n,p}$~\cite{magmoms}, axial-current matrix elements $\langle
p \vert A_\mu \vert p
\rangle$~\cite{BEK}, etc.

According to the Skyrme model, the proton contains many relativistic current
quarks, including $\bar s s$ pairs generated by the SU(3) rotations in
(\ref{four}).  The nucleon angular momentum is generated in this picture by the
slow rotation of the coherent meson cloud, so that if one decomposes the nucleon
helicity in the infinite-momentum frame:
\beq
\frac{1}{2} = \frac{1}{2} \sum_q \Delta q + \Delta G + L_Z~,
\label{six}
\eeq
where the $\Delta q(\Delta G)$ are the net contributions of the quark (gluon)
helicities, one predicts~\cite{BEK,EK}
\beq
\sum_q \Delta q = \Delta G = 0,~~L_Z = \frac{1}{2}~.
\label{seven}
\eeq
In the meson picture, $L_Z = \frac{1}{2}$ arises from the mixing of isospin and
conventional spin in the coherent cloud.  In the quark picture, it should be
interpreted statistically as an expectation value $\langle p\vert L_z \vert p
\rangle = \frac{1}{2}$, much as inside the Deuteron $\langle D \vert L_z \vert D
\rangle \simeq 0.06$ as a result of $D$-wave mixing.
In this picture, the fact that $\Delta \Sigma = \frac{1}{2} \sum_q \Delta q = 0$
is a consequence of the topology of the internal SU(2) (or SU(3)) flavour group,
and has nothing to do with the anomalous axial U(1) symmetry of
QCD~\cite{SV}.

The Skyrme model is not necessarily in conflict with the idea of constituent
quarks, and several models of chiral constituent quarks have been
proposed~\cite{chiconstq}, in
which $\vert U \rangle = \vert u \rangle + u \bar q q \rangle + \vert u G
\rangle +$ ... .  However, as yet none of these has been derived rigrously from
QCD.

As already recalled, the OZI rule~\cite{OZI} is to draw only connected
quark line
 diagrams.  This gains predictive power when it is further assumed that hadrons
 have only their na\"\i ve flavour compositions:  $\vert p \rangle = \vert uud
 \rangle, \vert \phi \rangle = \vert \bar s s \rangle$, etc...
However, it is known that OZI-forbidden processes such as $\phi \rightarrow
3\pi, f'_2(1520) \rightarrow 2\pi, J/\psi \rightarrow$ hadrons and $\psi'
\rightarrow J/\psi + \pi \pi$ do occur at levels $\lappeq 10^{-2}$.  In the
cases of $\phi$ and $f'_2$ decay, these violations are conventionally ascribed
to $\vert \bar u u + \bar d d \rangle$ admixtures in the meson wave functions,
whereas in the cases of $J/\psi$ and $\psi'$ decay they are ascribed to
pair-creation processes mediated by gluon exchanges.  Are all OZI-forbidden
processes restricted to the level $\lappeq 10^{-2}$, and can they all be
explained by meson mixing or pair creation?

\section{Prehistory}

There has long been some evidence that there may be $\bar s s$ pairs in the
nucleon.  The first example may have been the $\pi -N$ $\sigma$
term~\cite{sigma}:
\beq
\Sigma^{\pi N} \equiv \frac{1}{2} (m_u + m_d) \langle p \vert (\bar u u + \bar d
d) \vert p \rangle~,
\label{eight}
\eeq
which was first estimated using the Gell-Mann-Okubo mass formula and assuming
$\langle p \vert \bar s s \vert p \rangle = 0$, to obtain
\beq
\Sigma^{\pi N}_{\rm OZI} \simeq 25~{\rm MeV}~.
\label{nine}
\eeq
However, the experimental value (hedged about with qualifications associated
with the extrapolation from the Cheng-Dashen point, etc.), is estimated to
be~\cite{sigma}
\beq
\Sigma^{\pi N}_{\rm exp} \simeq 45~{\rm MeV}~.
\label{ten}
\eeq
The discrepancy between (\ref{nine}) and (\ref{ten}) corresponds to
\beq
y_N \equiv \frac{2 \langle p \vert \bar s \vert p \rangle}{\langle p \vert (\bar
u u + \bar d d) \vert p \rangle} \simeq 0.2~,
\label{eleven}
\eeq
with an uncertainty that may be $\pm 0.1$, whereas chiral symmetry and
the successes of the pseudoscalar-meson mass formulae (\ref{two}) suggest that
$y_\pi \sim$ few \%.  For comparison with the experimental 
value (\ref{eleven}), we
recall that a Skyrme calculation~\cite{Skyrmesigma} yields $y_N = 7/23$.

A second {\it a priori} example of OZI violation was the presence of $\bar s s$
pairs in the sea part of the proton wave function revealed by charm production
in {\it deep-inelastic neutrino scattering} on an unpolarized
target~\cite{NuTeV}. The
reactions $\FF + N \rightarrow \mu^\mp$ + charm + $X$ receive important
contributions from $s \rightarrow c$ transitions $\propto \cos^2\theta_c$, as
well as from $d \rightarrow c$ transitions $\propto \sin^2\theta_c$.  Several
experiments have found the need for an important $\bar s s$ contribution, and
the recent NuTeV analysis~\cite{NuTeV} is stable when extended from LO to
NLO QCD, yielding
the following ratio of integrals of parton densities:
\beq
\kappa \equiv
\frac{2\int^1_0dx (s + \bar s)}{\int^1_0 dx (u + \bar u + d + \bar d)} = 0.42 \pm
0.08~,
\label{twelve}
\eeq
for $Q^2 \simeq 16$~GeV$^2$~~\footnote{The result 
shows no strong $Q^2$ dependence.}. 
The $x$ distributions of the $s$ and $\bar s$ parton 
distributions appear similar to
each other, and comparable to those of 
the $\bar u u$ and $\bar d d$ sea components:
\beq
xs(x) \propto (1-x)^\beta~:~\beta = 8.5 \pm 0.7~,
\label{thirteen}
\eeq
and $-1.9  < \beta - \bar \beta < 1.0$~\cite{NuTeV}.  These can be
regarded as measurements of
an infinite tower of local twist-2 operator matrix elements:  $\langle N \vert
\bar s \gamma \partial^\mu s \vert N \rangle \not= 0$, which decrease relative
to the corresponding $\langle N \vert \bar q \gamma\gamma^nq \vert N\rangle$
because of the harder $x$ distribution of valence quarks.

The results (\ref{eleven},\ref{twelve},\ref{thirteen}) together imply that there
are many non-zero matrix elements $\langle N \vert \bar s(...) s \vert N
\rangle$,
though they may depend on the space-time properties~\cite{IoffeK}.

A first indication that the strange axial-current matrix element $\langle
p
\vert \bar s \gamma_\mu \gamma_5 s \vert p \rangle \equiv 2s_\mu \cdot 
\Delta s \not= 0$, where $s_\mu$ is the proton spin vector,
came from measurements in
{\it elastic $\FF$ p scattering} of the $\langle p \vert \bar s \gamma_\mu
\gamma_5 s \vert p \rangle$ matrix element~\cite{nup}, but this was not
noticed until after
the first EMC measurements~\cite{EMC} of {\it polarized deep-inelastic
structure functions}.

\section {The Strange Proton Spin}

As is well known~\cite{poldis}, polarized deep-inelastic electron or muon
scattering is
characterized by two spin-dependent structure functions $G_{1,2}(\nu,Q^2)$:
\beq
\frac{d^2\sigma^{\uparrow\downarrow}}{dQ^2d_\nu} -
\frac{d^2\sigma^{\uparrow\uparrow}}{dQ^2d_\nu} = \frac{4\pi \alpha^2}{Q^2E^2}
\left[ m_N(E+E'\cos \theta)G_1(\nu,Q^2) 
 -   Q^2G_2(\nu,Q^2) \right]~.
\label{fourteen}
\eeq
In the Bjorken scaling limit:  $x \equiv Q^2/2m_N\nu$ fixed and $Q^2 \rightarrow
\infty$, the na\"\i ve parton model predicts the scaling properties
\beq
m^2_N\nu G_1(\nu,Q^2) \rightarrow g_1(x)~,~~m_N\nu^2G_2(\nu,Q^2) \rightarrow
g_2(x)~,
\label{fifteen}
\eeq
where $g^p_1(x)$ has the following representation in terms of the different
helicities and flavours of partons:
\beq
g^p_1(x) = \frac{1}{2} \sum_q~e^2_q \left[q_\uparrow(x) - q_\downarrow(x) + \bar
q_\uparrow(x) -\bar q_\downarrow(x) \right] \equiv \frac{1}{2}
\sum_q~e^2_q\Delta q(x)~.
\label{sixteen}
\eeq
This expression can be compared with that for the unpolarized structure function
$F_2(x) = 2xF_1(x)$:
\beq
F_2(x) = \sum_q~e^2_qx \left[ q_\uparrow (x) + q_\downarrow (x) + \bar q_\uparrow
(x) + \bar q_\downarrow (x) \right]~.
\label{seventeen}
\eeq
The quantity measured directly is the polarization asymmetry
\beq
A_1(x,Q^2) \equiv
\frac{\sigma_{1/2} - \sigma_{3/2}}{\sigma_{1/2} + \sigma_{3/2}}
\buildrel{B_j} \over{\rightarrow}
\frac{\sum_q~{e^2_q} [q_\uparrow(x) - q_\downarrow(x) + \bar q_\uparrow(x) - \bar
q_\downarrow(x)]}
{\sum_q~e^2_q [q)_\uparrow(x) + q_\downarrow(x) + \bar q_\uparrow(x) + \bar
q_\downarrow(x)]}~,
\label{eighteen}
\eeq
or the related asymmetry $g_1(x,Q^2)/F_1(x,Q^2)$.

Much interest has focussed on the integrals
\beq
\Gamma^{p,n}_1(Q^2) \equiv \int^1_0dx~g^{p,u,d}_1(x,Q^2)~,
\label{nineteen}
\eeq
which have the following flavour compositions in the na\"\i ve parton model:
\beq
\Gamma^b_1 = \frac{1}{2} \left( \frac{4}{9} \Delta u + \frac{1}{9} \Delta d +
\frac{1}{9} \Delta s \right)~,~~\Gamma_1 = \frac{1}{2} \left( \frac{4}{9} \Delta
d + \frac{1}{9} \Delta u + \frac{1}{9} \Delta s \right)~,
\label{twenty}
\eeq
where the net quark helicities $\Delta q$ are related to axial-current matrix
elements:
\beq
\langle p \vert \bar q \gamma_\mu \gamma_5 q \vert \rangle \equiv 2s_\mu
\cdot \Delta q~,
\label{twentyone}
\eeq
where $s_\mu$ is the proton spin vector.  
Some combinations of the $\Delta q$ are
known from low-energy experiments.  
From neutron $\beta$ decay and isospin SU(2),
one has
\beq
\Delta u - \Delta d \equiv g_p = 1.2670~(35)
\label{twentytwo}
\eeq
and from a global fit to hyperon $\beta$ decays and flavour SU(3), one
has
\beq
\Delta u + \Delta d -2 \Delta s \equiv g_8 = 0.585~(25)
\label{twentythree}
\eeq
Using (\ref{twenty}) and (\ref{twentytwo}), we recover the sacred Bjorken sum
rule~\cite{Bj}
\beq
\Gamma^p_1 - \Gamma^n_1 = \frac{1}{6}g_A = \frac{1}{6} (\Delta u -\Delta d)
\label{twentyfour}
\eeq
It is amusing to recall that Bjorken famously dismissed this sum rule as
`worthless'~\cite{Bj}! However, it led to the prediction of scaling and
the formulation of
the parton idea, and is now recognized (with its calculable perturbative
corrections) as a crucial test of QCD, that the theory has passed with flying
colours.  Bjorken commented~\cite{Bj} that individual sum rules for the
proton and neutron
would depend on a model-dependent isotopic-scalar contribution, related in our
notation to $g_0 \equiv \Delta\Sigma \equiv \Delta u + \Delta d + \Delta s$.
The profane singlet sum rules~\cite{EJ} were derived assuming $\Delta s =
0$, motivated by
the idea that, even if the OZI rule was not valid for parton
distributions,
surely the sea quarks would be unpolarized.  As is well known, the data do not
support these na\"\i ve singlet sum rules, 
which is surely more interesting than if
they had turned out to be right.

The data on $\Gamma^{p,n}_1$ can be used to calculate $\Delta s$ and $\Delta
\Sigma = \Delta u + \Delta d + \Delta s$, with a key role being played by
perturbative QCD corrections~\cite{QCDcorr}:
\bea
\bigg[ 1 &-& \frac{\alpha_s(Q^2)}{\pi} - 1.0959 \left( \frac{\alpha_s(Q^2)}{\pi} \right)^2
- 4  \left( \frac{\alpha_s(Q^2)}{\pi} \right)^3 + \ldots \bigg] \Delta
\Sigma(Q^2) \nonumber \\
&=& \Gamma^{p,n}_1(Q^2) - \left( \pm \frac{1}{12} g_A + \frac{1}{36} g_8 \right)
\times \left[ 1 - \left( \frac{\alpha_s(Q^2)}{\pi} \right) - 3.8533
\left(\frac{\alpha_s(Q^2)}{\pi} \right)^2 \right. \nonumber \\
&-& \left. 20.2153 \left( \frac{\alpha_s(Q^2)}{\pi} \right)^3 - 130 \left(
\frac{\alpha_s(Q^2)}{\pi} \right)^4 - \ldots \right]~.
\label{twentysix}
\eea
The data on $\Gamma^{p,n}_1$ are highly consistent if these perturbative QCD
corrections are included,
yielding
\beq
\Delta n = 0.81 \pm 0.01 \pm ?~,~~ \Delta d = -0.45 \pm 0.01 \pm ?~,~~ 
\Delta s =
-0.10 \pm 0.01 \pm ?~,~~ \Delta\Sigma = 0.25 \pm 0.04 \pm ?
\label{twentyseven}
\eeq
at $Q^2 = {\rm 5}~GeV^2$~\cite{poldis}.  The unspecified second error in
(\ref{twentyseven})
reflects possible additional sources of error that are difficult to quantify,
including higher-twist corrections, the extrapolations of the measured structure
functions to low $x$, etc ...  However, there is clear {\it prima facie} 
evidence that $\Delta s \not= 0$.

The results (\ref{twentysix}, \ref{twentyseven}) can be compared with some
theoretical calculations.  For example, in the na\"\i ve Skyrme model with
$m_{u,d,s} \rightarrow 0$, one finds~\cite{BEK}
\beq
\Delta u = \frac{4}{7} g_A~,~~\Delta d = -\frac{3}{7}g_A~,~~\Delta s =
-\frac{1}{7} g_A~,~~\Delta\Sigma = 0~.
\label{twentyeight}
\eeq
As commented earlier, in this model the smallness of $\Delta\Sigma$ is a
consequence of the internal topology of the SU(3) flavour group.  The absolute
normalization of the $\Delta q$, namely $g_A$, is dependent on details of the
model such as higher-order interactions, but the ratios (\ref{twentyeight}) are
quite model-independent.  Substituting the experimental value $g_A = 1.26$ into
(\ref{twentyeight}), one finds
\beq
\Delta u = 0.71~,~~\Delta d = -0.54~,~~ \Delta s = -0.18~,
\label{twentynine}
\eeq 
which agree qualitatively with the experimental numbers (\ref{twentyseven}). 
Improvement may be possible if non-zero quark masses (particularly $m_s$) are
taken into account.

Several lattice calculations yield encouraging values of $\Delta \Sigma$:
\beq
\Delta\Sigma = 0.18 \pm 0.10~\cite{Latt1},~0.25 \pm 012~\cite{Latt2},
0.21 \pm 0.12~\cite{Latt3},
\label{thirty}
\eeq
but, here again, there are problems with $g_A$ and $g_8$:
\beq
g_A = -.907(20)~, g_8 = 0.484(18)~\cite{Latt3},
\label{thirtyone}
\eeq
that may indicate the importance of a correct unquenching of quark loops.  A
recent development has been a calculation~\cite{Lattangmom} of the total
quark angular momentum:
\beq
J_q \equiv \frac{1}{2} \Delta\Sigma + L_q = 0.30(7)~,
\label{thirtytwo}
\eeq
which indicates indirectly that the gluon contribution should be similar:
\beq
J_g = \Delta G + L_G \sim 0.2~.
\label{thirtythree}
\eeq
One may hope in the future for considerable refinement of the present generation
of lattice calculations:  the challenge may then be to 
understand the physical mechanisms underlying the results found.

The perturbative evolution of polarized structure functions is  well
understood:
\bea
g_1(x,t) &=& \frac{1}{2} \langle e^2\rangle \int^1_x \frac{dy}{y}\times
\left[ C^s_q \left( \frac{x}{y}, \alpha_s(t)\right)\Delta\Sigma(y,t) +
2N_fC_g\left( \frac{x}{y},\alpha_s(t)\right)\Delta G(y,t) \right.\nonumber \\
&+& \left.C^{NS}_q \left( \frac{x}{y},\alpha_s(t)\right) 
\Delta q^{NS}(y,t)\right]
\label{thirtyfour}
\eea
where $t \equiv \ln Q^2/\Lambda^2$, the coefficient 
functions $C^s_q$, etc., are all known to
${\cal O}(\alpha_s(t))$, and the scale-dependent 
parton distributions are controlled by
evolution equations characterized by splitting functions $P_{ij}$ that are also
known to ${\cal O}(\alpha_s(t))$. Thus complete 
calculations to NLO are available~\cite{NLO}. The individual
${\cal O}(\alpha_s(t))$ correction terms are renormalization-scheme 
dependent, but the
complete physical results are of course scheme-independent.

One of the issues arising at NLO is the possible impact of polarized
glue~\cite{polglue}. It is
known that $\Delta G \sim 1/\alpha_s$, which introduces an important ambiguity
into the specification of the polarized quark distribution: since different
possible definitions are related by
\beq
\Delta q_1(x,Q^2) = \Delta q_2(x,Q^2) + {\cal O} (\alpha_s) \Delta G (x,Q^2)
\label{thirtyfive}
\eeq
one finds an ${\cal O}(1)$ ambiguity $\delta (\Delta q ) = {\cal O}(1)$. 
One possible
prescription $(\overline{MS})$ is simply to define $\Delta q(x,Q^2)$ by the
structure function $g_1(x,q^2)$. Another (AB) is to define 
the $\Delta q$ so that
$a_0$ (like $a_3$ and $a_8$) is independent of $Q^2$, which implies that
\beq
\Delta q_{\overline{MS}} = \Delta q_{AB} - {\alpha_s\over 2\pi}~\Delta G
\label{thirtysix}
\eeq
To this order, one cannot distinguish between $\Delta G_{\overline{MS}}$ and
$\Delta G_{AB}$. All-orders fits to data should give the same result whatever
prescription is used. However, at finite order they well differ, which provides
one estimator for possible theoretical errors in the analysis. Typical AB fits
yield~\cite{MSfit}
\beq
\Delta G = 1.6 \pm 0.9 ~,
\label{thirtyseven}
\eeq
providing an indication that $\Delta G > 0$, but no more. 
A corresponding AB fit~\cite{MSfit}
yields
\beq
a_0 = 0.10 \pm 0.05 \pm 0.07^{+0.17}_{-0.11}~,
\label{thirtyeight}
\eeq
where the first error is experimental, the second is due to the low-$x$
extrapolation, and the third is associated with the fit. If $\Delta G$ is as
large as (\ref{thirtyseven}), it opens up the possibility that 
\beq
\Delta s_{AB} = \Delta s _{\overline{MS}} + {\alpha_s\over 2\pi}~\Delta G \simeq
0~,
\label{thirtynine}
\eeq
which might be thought to rescue the OZI rule~\cite{polglue}. It is
therefore of great
importance to try to measure $\Delta G$ directly. 

A first attempt was made in a search for a production asymmetry in $\vec p \vec
p \rightarrow \pi^0 + X$~\cite{FNAL}. There was no sign of a positive
signal, but the
theoretical interpretation is not very clean. A more recent attempt is via the
large-$p_T$ hadron-pair production asymmetry in photoproduction:$\vec\gamma\vec
p \rightarrow (h^+h^-)+X$. A negative asymmetry: $A = -0.28 \pm 0.12 \pm 0.02$
is found~\cite{HERMES}, which has the opposite sign from that expected
from $\gamma q
\rightarrow Gq$, and is consistent in magnitude and sign with many
polarized-gluon models. It therefore becomes important to confirm whether the
effect is non-zero, and we also await eagerly data 
from COMPASS~\cite{COMPASS}, from polarized beams at RHIC~\cite{polRHIC},
and from polarized beams at HERA~\cite{polHERA}.

\section{Hadronic Probes of Hidden Strangeness}

According to the na\"\i ve OZI rule~\cite{OZI}, if $A$, $B$ and $C$ are
non-strange
hadrons, then
\beq
Z_{ABC} \equiv {\sqrt{2} {\cal M} (AB\rightarrow C+\bar ss)\over {\cal
M}(AB\rightarrow C+\bar uu) + {\cal M} (AB\rightarrow C + \bar dd)} = 0
\label{fourty}
\eeq
In this case, the production of a predominantly $\bar ss$ meson such as $\phi$
or $f^\prime_2(1520)$ would be due to a departure $\delta = \theta - \theta_i$
from ideal mixing~\cite{EGK,EKKS1,EKKS2}, and, e.g.,
\beq
{{\cal M}(AB\rightarrow C\phi)\over{\cal M}(AB\rightarrow C\omega)} =
{Z_{ABC}+\tan\delta\over 1 - Z_{ABC} \tan\delta}
\label{fourtyone}
\eeq
In the case of the $\phi$ and $f^\prime_2$, squared-mass formulae suggest
\beq
\tan^2\delta_\phi \simeq 42 \times 10^{-3}~,~~~\tan^2\delta_{f^\prime_2} \simeq
16\times 10^{-3}
\label{fourtytwo}
\eeq
and in the latter case one may also estimate from the decay
$f^\prime_2\rightarrow \pi\pi$ that $\tan^2\delta_{f^\prime_2} = (2.6\pm
0.5)\times 10^{-3}$. There seem to be no particular problems for the OZI rule
provided by $\phi$ production in $\pi N$ collisions~\cite{EKKS2}:
\beq
R_{\pi N} \equiv {\sigma(\pi N\rightarrow \phi X)\over \sigma (\pi N \rightarrow
\omega X)} = (3.3\pm 0.3)\times 10^{-3}
\label{fourtythree}
\eeq
on average, making no phase-space corrections, whilst in $NN$
collisions~\cite{EKKS2}:
\beq
R_{NN} \equiv {\sigma(NN\rightarrow\phi X)\over \sigma(NN\rightarrow\omega X)} =
(14.7 \pm 1.5)\times 10^{-3} \Rightarrow Z_{NN} = (8.2 \pm 0.7) \%
\label{fourtyfour}
\eeq
and in $\bar pp$ annihilation in flight:
\beq
R_{\bar pp} = (11.3 \pm 1.4)\times 10^{-3} \Rightarrow Z_{\bar pp} = (5.0 \pm
0.6) \%
\label{fourtyfive}
\eeq
which are not particularly dramatic.

\begin{figure}[h]
\hglue3.5cm
\epsfig{figure=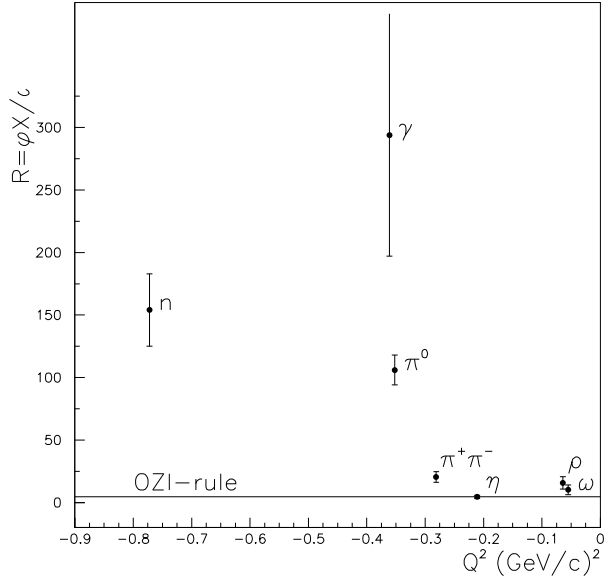,width=8cm}
\caption{\label{fig:ozi} {\it Compilation of LEAR data testing the
OZI rule in different $\bar p p$ annihilation channels.}}
\end{figure}

In this context, some of the data from $\bar pp$ annihilation at rest at LEAR
shown in Fig.~\ref{fig:ozi} came as a great shock,
especially~\cite{LEARgamma}
\beq
R_\gamma \equiv {\sigma(\bar pp\rightarrow\phi\gamma )\over \sigma(\bar
pp\rightarrow\omega\gamma)} = (294 \pm 97)\times 10^{-3}
\label{fourtysix}
\eeq
and~\cite{LEARpi}
\beq
R_{\pi^0} \equiv {\sigma(\bar pp\rightarrow\phi\pi^0)\over \sigma(\bar
pp\rightarrow\omega\pi^0)} = \left\{\matrix{(106\pm 12)\times 10^{-3} &{\rm
in~LH}_2\cr
(114\pm 24)\times 10^{-3} &~{\rm in ~H ~gas} \hfill}\right.
\label{fourtyseven}
\eeq
The $\phi$ production rates exhibited no universal factor, as might be expected
in a mixing model (or in a shake-out mechanism - see later), and were sometimes
strongly dependent on the initial state~\cite{LEARstate}:
\beq
B(\bar pp\rightarrow\phi\pi^0)\bigg\vert_{^3S_1} = (4.0\pm 0.8)\times 10^{-4}~,
~~~
B(\bar pp\rightarrow\phi\pi^0)\bigg\vert_{^1P_1} < 0.3\times 10^{-4}
\label{fourtyeight}
\eeq
To add to the mystery, there were some $\bar pp$ annihilation channels where no
large effect was observed~\cite{LEARno,LEARgamma}:
\beq
R_\eta = (4.6\pm 1.3)\times 10^{-3}~, ~~~ R_\omega = (19\pm 7) \times 10^{-3}~,~~~
R_\rho = (6.3 \pm 1.6)\times 10^{-3}
\label{fourtynine}
\eeq
The interpretation we propose~\cite{EKKS1,EKKS2} is that the proton wave
function contains
polarized $\bar ss$ pairs.

\begin{figure}[h]
\hglue4.0cm
\epsfig{figure=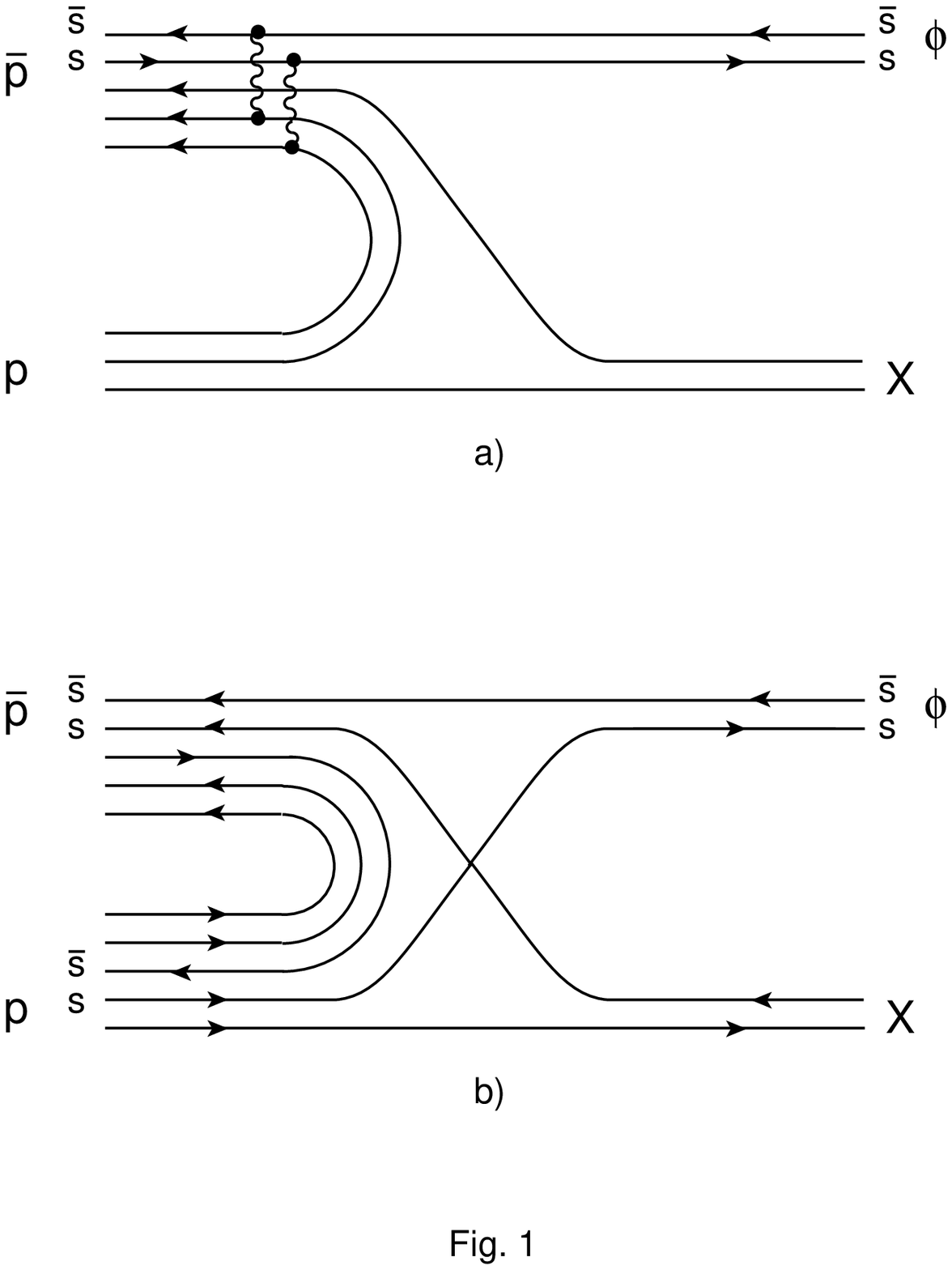,width=6cm}
\caption{\label{fig:quark} {\it Shake-out and rearrangement quark-line
diagrams that could contribute to $\phi$ or $f'$ production in
$\bar p p$ annihilation if the proton wave functions contains $\bar s s$
pairs~\cite{EKKS1,EKKS2}.}}
\end{figure}

In general, if there is an $\bar ss$ component in the Fock-space decomposition
of the proton wave function:
\beq
\vert p\rangle = x \sum_X \vert uud X \rangle + Z \sum_X \vert uud \bar ss X
\rangle
\label{fifty}
\eeq
where the remnant $X$ may contain gluons and light $\bar qq$ pairs, the na\"\i
ve OZI rule may be evaded by two new classes of connected quark-line diagrams,
shake-out and rearrangement~\cite{EKKS1,EKKS2} as illustrated in
Fig.~\ref{fig:quark}. The former yields an amplitude
\beq
{\cal M}_{SO}(\bar pp\rightarrow \bar ss + X) \simeq 2 Re (x^*z) P(\bar ss)
\label{fiftyone}
\eeq
where $P(\bar ss)$ is a projection factor that depends primarily on the final
state: $\bar ss = \phi, f^\prime_2,\ldots$. Rearrangement yields an amplitude
\beq
{\cal M}_R (\bar pp\rightarrow \bar ss + X) \simeq \vert Z\vert^2 T (\bar ss)
\label{fiftytwo}
\eeq
where $T(\bar ss)$  is a projection factor that may well depend on the initial
state as well as the final state.


There are infinitely many possibilities for the 
quantum numbers of the $\bar ss$
pairs in (\ref{fifty}). Assuming that
\beq
\vert p \rangle_{{1\over 2}}~ \ni ~\vert uud 
\rangle_{{1\over 2}} \otimes
\vert \bar ss \rangle~,
\label{fiftythree}
\eeq 
the simplest possibilities are those shown in Table~1. The first two
of
these
are disfavoured by data on $\eta$ production and by the non-universality of
$\phi$ production. We favour~\cite{EKKS2} a $^3P_0$ state for the $\bar
ss$, as in the vacuum
(\ref{one}). The triplet spin state should be antiparallel to the proton spin, 
as suggested by the polarized structure function data.

\begin{table}[hbt]
\label{tab:jpc}
\begin{tabular*}{\textwidth}{@{}l@{\extracolsep{\fill}}rrrrr}
\hline
S & L & j & $J^{PC}$ & State\\
\hline
0 & 0 & 1 & $0^{-+}$ &$^1S_0$ `$\eta$'\\
1 & 0 & 1 & $1^{--}$ &$^3S_1$ `$\phi$'\\
1 & 1 & 0 & $0^{++}$ &$^3P_0$ \\
1 & 1 & 0 & $1^{++}$ &$^3P_1$ \\
0 & 1 & 0 & $1^{+-}$ &$^1P_1$ \\
\hline
\end{tabular*}                         
\vspace*{0.2cm}   
\caption{{\it Possible
quantum numbers of the $\bar s s$ cluster in the nucleon.
We denote by $\vec{S}$ and $\vec{L}$ the
total spin and orbital angular momentum of the $\bar s s$ pair,
$\vec{J} \equiv \vec{L}+\vec{S}$, and
the relative angular momentum between
the $\bar s s$ and $uud$ clusters is $\vec{j}$.}}
\end{table}

\begin{figure}
\hglue3.0cm
\epsfig{figure=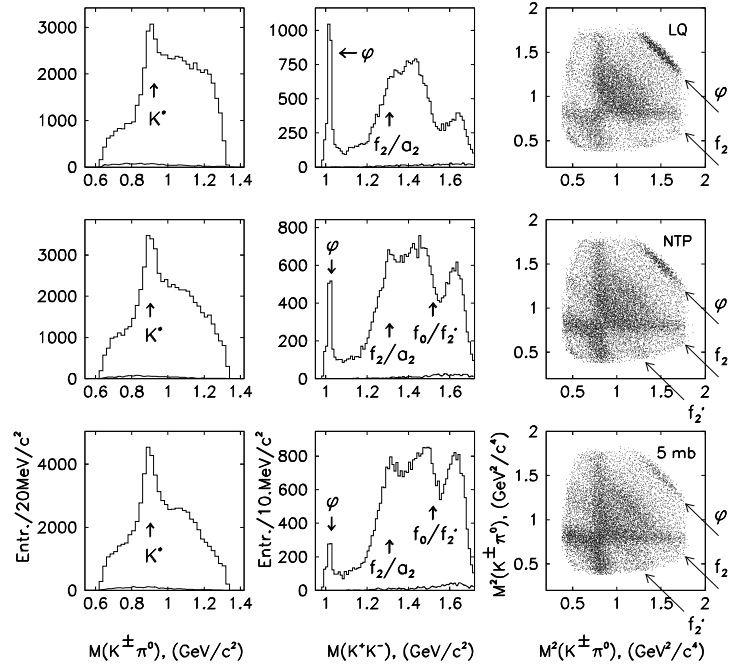,width=10cm}
\caption{\label{fig:dal} {\it The $K^\pm \pi^0$, $K^+ K^-$ mass and Dalitz
distributions in $\bar p p \rightarrow K^+ K^- \pi^0$: the
top (bottom) row of plots have more $S$-($P$-)wave annihilations, and the
$\phi$ ($f'$) is more visible.}}
\end{figure}

In such a picture, shake-out would yield predominantly $K^+K^-$ and $K^0\bar
K^0$ pairs in a relative $S$ wave: we have argued~\cite{EKKS2} that this
is consistent with
LEAR data. This picture also predicts that the $\phi$ should be produced more
strongly from $^3S_1$ $\bar pp$ initial states, whereas the $f^\prime_2$ should
be produced more from $^3P_J$ initial states. The former is consistent with the
previous data (\ref{fourtyseven}). What do more recent data indicate? The
enhancement (\ref{fourtyseven}) of $\phi$ production from the $^3S_1$ initial
state has now been confirmed with about 100 times more
statistics~\cite{LEARconfirm}:
\beq
B(\bar pp\rightarrow\phi\pi^0)\vert_{^3S_1} = (7.57\pm 0.62)\times 10^{-4}~, ~~~
B(\bar pp\rightarrow\phi\pi^0)\vert_{^1P_1} < 0.5\times 10^{-4},
\label{fiftyfour}
\eeq
as seen in Fig.~\ref{fig:dal}
whereas there is no similar trend for $B(\bar pp\rightarrow\omega\pi^0)$.
It has also
been observed that $\sigma(\bar np\rightarrow\phi\pi^+)$ decreases as energy
increases, similarly to the $S$-wave annihilation fraction, but there is no
similar trend for $\sigma(\bar np\rightarrow\omega\pi^+)$. The importance of
$\phi$ production from the $S$ wave is supported by the recent measurement
of~\cite{ppphi}
\beq
R_{pp} = {\sigma(pp\rightarrow pp\phi)\over \sigma(pp\rightarrow pp\omega)} =
(3.7\pm 0.7^{+1.2}_{-0.9}) \times 10^{-3}
\label{fiftyfive}
\eeq
at an energy 83 MeV above the $\phi$ production threshold. The phase-space
corrrection to (\ref{fiftyfive}) would be at least a factor of 10: in fact,
there are indications that $\omega$ production may be from a mixture of $S$ and
$P$ waves. There are also indications that $f^\prime_2$ production may be
enhanced in the $P$-wave initial state~\cite{LEARconfirm}:
\beq
R(f^\prime_2 \pi^0/f_2\pi^0)\vert_S = (47\pm 14)\times 10^{-3}~, ~~~
R(f^\prime_2 \pi^0/f_2\pi^0)\vert_P = (149\pm 20)\times 10^{-3}
\label{fiftysix}
\eeq
as also seen in Fig.~\ref{fig:dal}.
According to this picture, $\eta$ production should also be enhanced in
spin-singlet initial states, which is supported by the data~\cite{pneta} 
\beq
R_\eta \equiv {\sigma(np\rightarrow np\eta)\over \sigma (pp\rightarrow pp\eta )}
= {1\over 4}~ (1+\vert f_0/f_1\vert^2)
\label{fiftyseven}
\eeq
where $f_i$ denotes the amplitude for the isospin = spin = i initial state at
threshold. The measured value $R_3 \simeq 6.5$ suggests that $\vert
f_0/f_1\vert^2 \simeq 25$.

\begin{figure}
\hglue3.8cm
\epsfig{figure=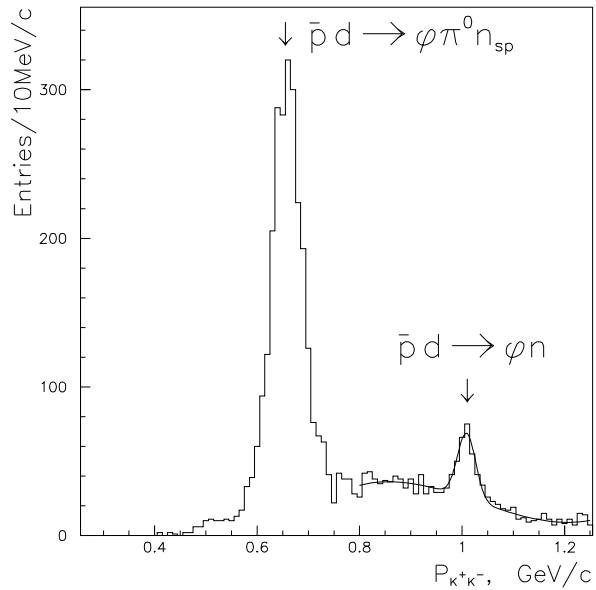,width=8cm}
\caption{\label{fig:ponte} {\it The momentum distribution of $K^+ K^-$
pairs in $\bar p D \rightarrow \phi + \dots$ shows clear peaks
corresponding to $\bar p D \rightarrow \phi \pi^0 n$ and the
Pontecorvo reaction $\bar p D \rightarrow \phi n$.}}
\end{figure}

An interesting recent development has been the observation of copious $\phi$
production in the Pontecorvo reaction~\cite{Pontecorvo}, shown in
Fig.~\ref{fig:ponte}:
\beq
R(\bar pd\rightarrow\phi n / \bar pd\rightarrow\omega n) = (154\pm 29)\times
10^{-3}
\label{fiftyeight}
\eeq
which is expected to be dominated by $S$-wave annihilation. On the other hand,
it has also been measured that~\cite{otherPont}
\beq
R(\bar pd\rightarrow K\Sigma / \bar pd\rightarrow K\Lambda ) = (0.92\pm 0.15)
\label{fiftynine}
\eeq
whereas a two-step model predicted a ratio of 0.012.

There have recently been many calculations of two-step contribution to
$\bar
pp\rightarrow \phi\pi$ in particular, including three- as well as two-particle
intermediate states~\cite{Locher}. Individual intermediate states make
calculable
contributions to the imaginary part of the annihilation amplitude, but their
relative signs are not known, and not always their spin decompositions, either.
With suitable choices and estimates of the real parts, the data on $\sigma (\bar
pp\rightarrow \phi \pi^0)\vert_{^3S_1}$, can be fit. However, some challenges
remain. Can the two-step prediction be made more definite? Can good predictions
be made for other partial waves? Can the apparent anticorrelation with $K^*K$
yields be understood? or the energy dependences of $\phi\pi$ and $K^*K$ final
states? Can the data on $f^\prime_2\pi^0$ production and the Pontercorvo
reaction be understood? Can the apparent OZI violation be correlated with other
observables, as we discuss next?

\section{Further Tests in $\Lambda$ Production}

The total cross sction and angular distribution for $\bar pp\rightarrow \bar
\Lambda\Lambda$ were measured in the PS 185 experiment at LEAR, and in
particular the $\bar\Lambda\Lambda$ spin correlation was measured. It was
found~\cite{PS185}
that the spin-triplet state dominated over the spin-singlet state by about
2 orders of
magnitude. This triplet dominance could be accommodated in meson-exchange models
by suitable tuning of the $K$ and $K^*$ couplings. On the other hand, triplet
dominance is a natural prediction of gluon-exchange models, and also of our
$^3P_0$ $\bar ss$ model~\cite{AEK}. One way to discriminate between models
is to use a
polarized beam and measure the depolarization $D_{nn}$ (i.e., the polarization
transfer to the $\Lambda$)~\cite{AEK}. Polarized-gluon models would
predict positive
correlations between the $p, g, s$ and $\Lambda$ spins, so that $D_{nn} > 0$,
whereas meson-exchange models predict $D_{nn} < 0$. The polarized-strangeness
model predicts an anti-correlation of the $p$ and $s$ spins, and hence $D_{nn} <
0$. Data with a polarized $p$ beam have been taken, and we await the analysis
with interest. They may be able to distinguish between polarized-gluon and
polarized-strangeness models. In the mean time, it is interesting that DISTO has
recently measured~\cite{DISTO} $D_{nn} < 0$ in the reaction $\vec
pp\rightarrow\Lambda K^+p$,
in agreement with the polarized-strangeness and meson-exchange models.


\begin{figure}
\hglue3.0cm
\epsfig{figure=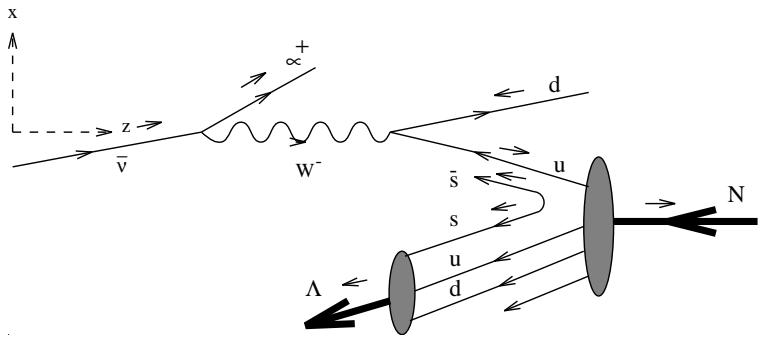,width=10cm}
\caption{\label{fig:nu1} {\it In deep-inelastic scattering with a
polarized beam ($\bar \nu$ in the drawing), the polarized
intermediate boson ($W$ in the drawing) selects preferentially one
particular polarization of struck quark ($u$ in the drawing), which may in
turn select preferentially one
polarization of the spectator $\bar s s$ pair, which may be carried over
to the spin of a $\Lambda$ in the target fragmentation region.}}
\end{figure}

\begin{figure}
\hglue3.0cm
\epsfig{figure=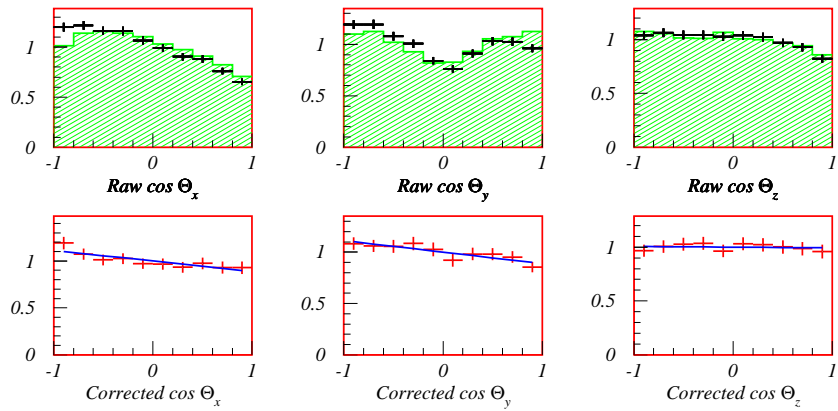,width=11cm}
\caption{\label{fig:lachaud7} {\it Data on
deep-inelastic $\nu N$ scattering~\cite{NOMAD} indicate negative $\Lambda$
longitudinal polarization (left panel), as well as transverse polarization
in the
scattering plane (centre panel), but not outside it (right panel).}}
\end{figure}

Another potential test is $\Lambda$ polarization in deep-inelastic
scattering~\cite{EKK}, as illustrated in Fig.~\ref{fig:nu1}. 
Here  the key idea is that when a polarized lepton $(\bar \nu , e, \mu$ or
$\nu$)  scatters via a polarized boson $(W^*$ or $\gamma^*$), it selected
preferentially a particular polarization of the struck quark ($u$ or $d$) in the
nucleon target, even if the latter is unpolarized. The next suggestion is that
the target `remembers' the spin removed, e.g., $\bar p> \rightarrow
u^\uparrow +
(u^\uparrow d^\downarrow (\bar ss)^\downarrow)$. The polarization of the $s(\bar
s)$ may then be retained by a $\Lambda(\bar \Lambda )$ in the target fragmentation
region. This prediction~\cite{EKK,Ketal} was supported by early data on
$\Lambda$ polarization
in deep-inelastic $\bar \nu N$ data. Recent NOMAD data~\cite{NOMAD} on
deep-inelastic $\nu N$
scattering confirm this prediction with greatly increased statistics,
as seen in Fig.~\ref{fig:lachaud7}:
the $\Lambda$ polarization in the direction of the exchanged $W$ is
$- 0.16 \pm 0.03 \pm 0.02$. The
measurement of $\Lambda$ polarization in the target fragmentation region is also
in the physics programmes of HERMES in deep-inelastic $eN$ scattering and of
COMPASS in deep-inelastic $\mu N$ scattering.

\section{Strangeness Matters}

The presence or absence of hidden strangeness in the proton is relevant to many
other experiments in other areas of physics. Here I just mention just one
example~\cite{EFR}: the search for cold dark matter~\cite{GW}. The idea
is that a
massive
non-relativistic neutral particle $\chi$ may strike a target nucleus depositing
a detectable amount of recoil energy $\Delta E \sim m_\chi v^2_\chi/2 \sim$ keV.
The scattering cross section has in general both spin-dependent and
spin-independent pieces. The former may be written as~\cite{EFO}
\beq
\sigma_{spin} = {32\over\pi}~ G^2_F~\hat m^2_\chi \Lambda^2~ J(J+1)
\label{sixty}
\eeq
where $\hat m_\chi$ is the reduced mass of the relic particle, $J$ is the spin
of the nucleus, and
\beq
\Lambda = {1\over J}~(a_p \langle S_p \rangle + a_n \langle S_n \rangle)
\label{sixtyone}
\eeq
where
\beq
a_p = \sum_{q=u,d,s} \hat\alpha_q ~~\Delta q
\label{sixtytwo}
\eeq
(and similarly for $a_n$) with the coefficients $\hat\alpha_q$ depending on the
details of the supersymmetric model, and the $\Delta q$ being the familiar quark
contributions to the proton spin measured by EMC et al. Likewise, 
the spin-independent
part of the cross section can be written as
\beq
\sigma_{scalar} = {4 \hat m^2_\chi\over\pi}~~\bigg[ Z f_p + (A-Z) f_n \bigg]^2
\label{sixtythree}
\eeq
where $Z$ and $A$ are the charge and atomic number of the nucleus, and
\beq
f_p = m_p~ \sum_{q=u,d,s}~ f_{T_q}~{\alpha_q \over m_q} + {2\over 27}~ f_{T_G}
\sum_{q=c,b,t}~ {\alpha_q\over m_q}
\label{sixtyfour}
\eeq
(and similarly for $f_n$) where the coefficients $\alpha_q$ again depend on the
details of the supersymmetric model, and
\beq
m_p~f_{T_q} \equiv \langle p\vert m_q \bar qq \vert p \rangle~, ~~~ f_{T_G} = 1 -
\sum_{q=u,d,s} f_{T_q}
\label{sixtyfive}
\eeq
We depend on measurements of the $\pi N$ $\sigma$ term for our knowledge of the
$f_{T_q}$, and $\langle p\vert\bar ss \vert p\rangle$ plays a key
role~\cite{EFO}.

\begin{figure}
\vspace*{-0.8in}
\hspace*{-.1in}
\begin{minipage}{10in}
\epsfig{file=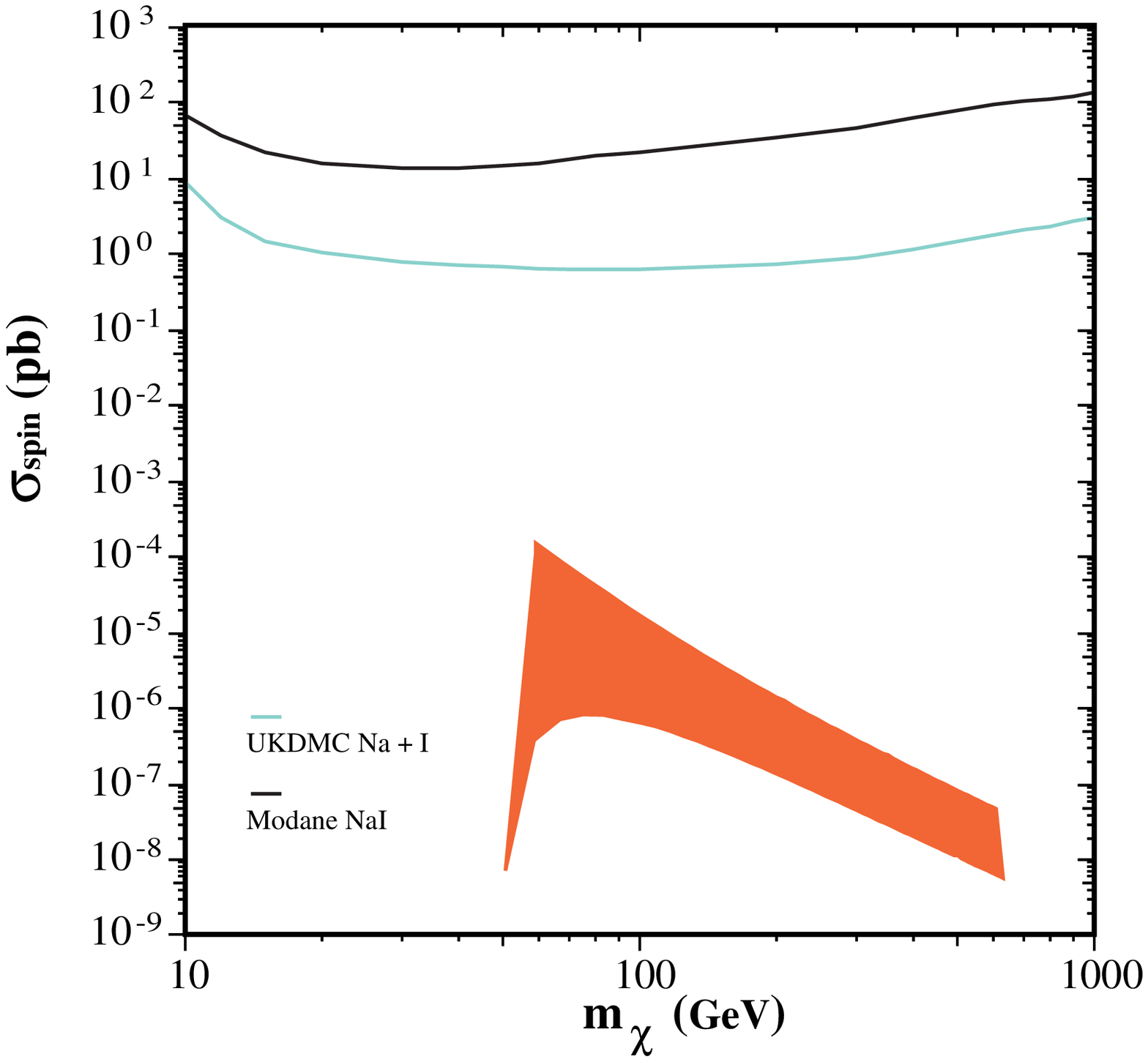,width=8cm} 
\hspace*{-.20in}\epsfig{file=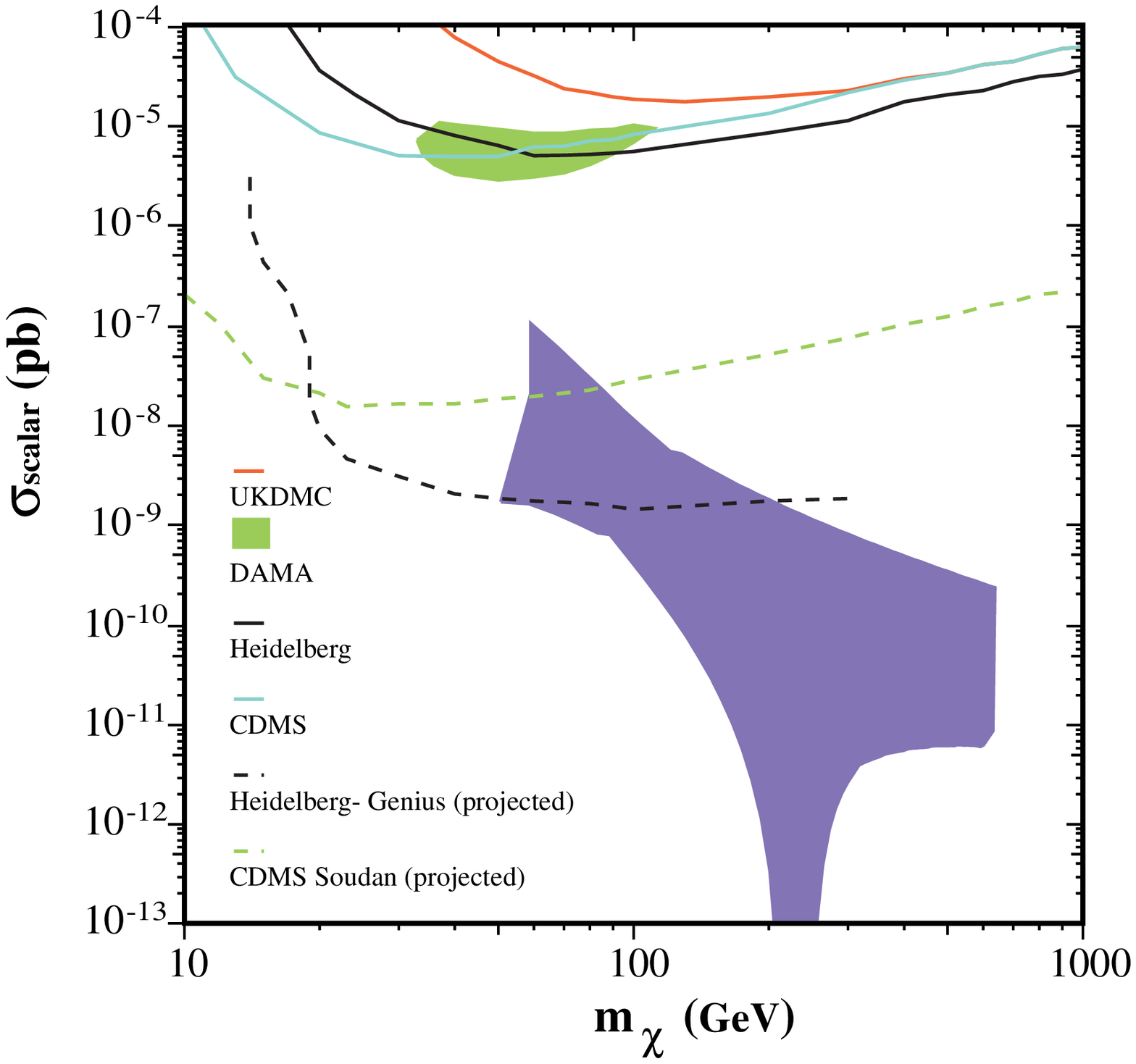,width=8cm} \hfill
\end{minipage}
\vskip-0.11in
\caption{\label{fig:fig4} {\it Estimates~\cite{EFO} (a) of the
spin-dependent dark
matter scattering
rate in the minimal supersymetric extension of the Standard Model,
compared with experimental upper limits, and (b) of
the corresponding
spin-independent dark matter scattering rate,
compared with upper limits and the measurement reported in~\cite{DAMA}.}}
\end{figure}

Fig.~\ref{fig:fig4}(a) compares a prediction for supersymmetric cold dark
matter (in a
favoured model) with experimental upper limits on the spin-dependent cross
section, assuming that our galactic halo is dominated by supersymmetric relic
particles~\cite{EFO}. A similar comparison for the spin-independent cross
section is shown
in Fig.~\ref{fig:fig4}(b). In this case, there is one experiment that
reports possible
evidence for a signal~\cite{DAMA}. The uncertainties in the strangeness
content of the
proton are not sufficient to explain the discrepancy with our prediction.
Perhaps our supersymmetric model is wrong? It would be good to see the reported
detection confirmed, but this has not happened yet~\cite{CDMS}.

\section{Issues for the Future}

Many new experiments may cast light on the `strange' nucleon wave 
function.
These include $\pi N$ scattering for the $\sigma$ term, polarized $eN$ structure
functions and final-state asymmetries, $\phi$ and $f'$ production 
in hadro-, photo- and electro-production~\cite{electrophi}, $\Lambda$
polarization measurements in
polarized $eN, \mu N$ and $\nu N$ scattering, charm production asymmetries in
polarized $\mu N$ scattering, further data on OZI `violations' in $\bar
pp$
annihilation, $\pi p / pp$ scattering, electro- and photoproduction. Low-energy
experiments on parity violation in atomic physics~\cite{CEF} and $eN$
scattering will also be useful.

Meanwhile, there are many theoretical challenges. Can the different models make
quantitative  predictions? Are different theoretical approaches really in
conflict, or is there any sense in which they are different languages for
describing the same thing? We need to understand better the dialectics between
$\Delta S$ and $\Delta G$, between constituent quarks and chiral symmetry,
between intrinsic $\bar ss$ models~\cite{EKKS1,EKKS2} and the two-step
approach~\cite{Locher}. Surely none of
these religions has a monopoly of truth!

\vspace{1cm}
\noindent
{\bf Acknowledgements}:
It is a pleasure to thank my collaborators on topics discussed here,
including Mary Alberg, Marek Karliner, Dima Kharzeev, Aram
Kotzinian and particularly Misha Sapozhnikov.

\end{document}